# Automated Lesion Segmentation of Stroke MRI Using nnU-Net: A Comprehensive Exploration of Acute and Chronic Lesions

**Short title:** Automated Stroke Lesion Segmentation

Tammar. Truzman[1]*, Matthew A. Lambon Ralph[1] & Ajay D. Halai[1]

**Affiliation:**
[1]MRC Cognition and Brain Sciences Unit,
University of Cambridge,
15 Chaucer Road,
Cambridge CB2 7EF, UK

***Corresponding author:**
Tammar Truzman
15 Chaucer Road,
Cambridge CB2 7EF, UK
*Tammar.truzman@mrc-cbu.cam.ac.uk

**Highlights**:

- nnU-Net models achieved lesion-segmentation performance at human expert levels for both acute and chronic stroke under specific conditions.
- DWI consistently outperformed FLAIR in acute stroke lesion segmentation in the data available.
- Performance gains diminished beyond a few hundred cases with chronic T1-weighted datasets.
- Lesion laterality did not impact segmentation accuracy or generalisability.
- Lesion volume was a key performance determinant: while generalisation required a broad lesion volume distribution during training, very small lesions remained challenging to segment.
- Models trained on high-quality MR scans generalised well to lower-quality data, but not vice versa.

## Abstract

Accurate and generalisable segmentation of stroke lesions from magnetic resonance imaging (MRI) is essential for advancing clinical research, prognostic modelling, and personalised interventions. Although artificial intelligence, including deep learning, has substantially improved automated lesion delineation, many existing models are optimised for narrow imaging contexts, are not open-access, and generalise poorly to independent datasets, modalities and stroke stages. Here, we systematically evaluated stroke lesion segmentation using the nnU-Net framework across multiple heterogeneous, publicly available MRI datasets, spanning acute and chronic stroke. Models were trained and tested on diffusion weighted imaging (DWI), fluid-attenuated inversion recovery (FLAIR), and T1-weighted MRI, and were evaluated on fully independent datasets. Across acute and chronic stroke, the best-performing models achieved robust generalisation, with segmentation accuracy commensurate with inter-rater reliability for manual annotations. Performance varied systematically as a function of imaging modality and training data characteristics. For acute stroke, models trained on DWI consistently outperformed those trained on FLAIR, as expected given that the reference lesion masks were delineated on DWI. In chronic stroke, increasing training set size showed modest improvements beyond a few hundred cases. Lesion volume emerged as an important determinant of accuracy: smaller lesions were consistently harder to segment, and models trained on restricted volume ranges generalised poorly. Furthermore, MRI image quality was linked with model generalisability: models trained on lower-quality scans transferred poorly, whereas those trained on higher-quality data generalised well even to noisier images. Inspection of failed cases revealed that discrepancies between model prediction and reference masks were often attributable to inconsistencies or limitations in manual annotations, highlighting how variability in current gold standard labels constrains performance. Together, these findings demonstrate that automated lesion segmentation can achieve comparable levels to humans, while identifying key factors that govern generalisability. By leveraging open datasets, open-source algorithms, and code, this work provided practical guidance for developing robust, transparent, and clinically relevant stroke lesion segmentation tools.

## Introduction

Stroke remains a leading cause of death and long-term adult disability worldwide, with its prevalence and societal impact increasing in aging populations [1]. Brain damage following stroke is highly heterogeneous, and different neuroimaging modalities are informative at different stages of the disease; for example, perfusion-based imaging in the acute phase and structural imaging in the chronic phase. Rapid and accurate identification of stroke lesions is therefore essential not only for clinical diagnosis, prognosis, and disease monitoring [2,3], but also for advancing clinical research in stroke and its recovery. Lesion-symptom mapping has been central to cognitive neurology, neuropsychology and aphasiology since classical post-mortem studies by Broca and Wernicke, and contemporary neuroimaging now enables non-invasive, high-resolution three-dimensional reconstruction of damaged tissue. This has facilitated the development of quantitative approaches such as voxel-based lesion symptom mapping (VLSM) [4] and multivariate lesion-symptom methods [5,6]. Furthermore, precise lesion identification can support participant stratification in targeted therapeutic trials [7] and the identification of imaging biomarkers predictive of post-stroke recovery trajectories [8–12].

Despite its importance, accurate lesion identification remains challenging due to substantial variability in lesion appearance, imaging protocols, and tissue characteristics across patients and centres [13]. Manual lesion tracing is considered the gold standard but is time-consuming, requires specialised expertise, and is subject to inter- and intra-rater variability [14,15]. Even under standardised protocols, inter-rater reliability for chronic lesion tracing is approximately .72 and intra-rater reliability .84 [16,17], motivating the development of automated methods to improve efficiency, objectivity and reproducibility [18].

In the first wave of development, several automated or semi-automated approaches were proposed [19–22]; however, generalisation to external datasets was limited, particularly when applied to data from diverse sites, scanners, and stroke populations. Ito et al., [14] compared four automated methods and found that the best algorithm (using LINDA; [19]) achieved a median Dice coefficient (proportion of overlap between ground truth and predicted lesion) of .5 on stroke data collected worldwide. This performance falls below reported inter-rater reliability (cf. .72), substantially limiting its utility for clinical and research application.

More recently, advances in artificial intelligence, particularly deep learning, have led to substantial improvements in automated stroke lesion segmentation. However, reported model accuracy must be interpreted in light of several critical considerations. First, lesion appearance evolves markedly over time following stroke. Acute lesions are dominated by hypoperfusion and cytotoxic oedema, whereas chronic lesions reflect tissue loss, cavitation and gliosis. This temporal evolution directly determines which neuroimaging modalities are most informative for lesion detection, which, in turn, influences algorithms performance [23,24]. Consequently, some approaches have developed models tailored to specific post-stroke phases [25], while others have sought to build models that generalise across time points [26]. Second, the heterogeneous nature of stroke lesions often necessitates multi-modal imaging, introducing a trade-off between models optimised for specific modalities (or sets) [25] and modality-invariant models intended to be flexible across diverse protocols [26]. Increased flexibility may improve applicability but can come at the cost of reduced segmentation accuracy. Finally, evaluation strategies critically influence conclusions about generalisability. While random training, validation and test sets splits within a dataset are common, they may overstate real-world performance. In contrast, validation on independent datasets collected at different sites and scanners provides a stringent and clinically meaningful test. This distinction is evident in prior work: Ahmed et al. [25] reported high Dice coefficient (.82) using held-out cases from ATLASv2 [17], whereas Chalcroft et al. [26] demonstrated marked performance degradation when models were required to generalise across datasets, modalities, or stroke stages.

In summary, while recent studies demonstrate considerable promise, there remains a clear need for models that can reliably segment both acute and chronic lesions with uni- or multi-modal imaging, and that are rigorously evaluated on independent datasets. Widespread adoption of automated approaches will depend not only on high performance, but also on transparency, open availability of models and code, and feasibility of deployment by users without specialist expertise in computer science.

In the current study, we utilise a streamlined and robust implementation of the original U-Net convolutional neural network architecture (CNN; Ronneberger et al., 2015), specifically the “no-new-U-Net” (nnU-Net; [28]). nnU-Net is a fully automated, self-configuring framework that adapts network architecture and training parameters to the dataset at hand. Its strength lies in the end-to-end automation of the segmentation pipeline - from pre-processing, model configuration, training, inference and post-processing, while retaining flexibility for expert users where needed. Variants of nnU-Net have consistently achieved strong performance across diverse clinical imaging tasks [24,29]. Using nnU-Net, we systematically evaluated the performance and generalisability of automated stroke lesion segmentation across multiple openly available datasets containing manually delineated acute and chronic stroke lesions. Our primary aim was to develop models capable of reliably segmenting lesions across both time points, achieving performance approaching reported levels of human inter-rater agreement.

Beyond establishing strong baseline performance, the availability of multiple heterogeneous datasets allowed us to investigate key determinants of model generalisation. For acute lesion segmentation, we tested whether multimodal models would outperform unimodal models by integrating complementary information about tissue injury. For chronic lesion segmentation, we examined the effects of: (i) training dataset size; (ii) lesion spatial distribution; (iii) lesion volume distribution; and (iv) MRI scan quality in both training and test data. Consistent with general deep-learning principles, we hypothesised that larger and more diverse training datasets would improve generalisation.

All trained models, code and documentation are openly shared to promote transparency, reproducibility, and community-wide development, and are available on GitHub at https://github.com/AjayHalai/Lesion-Segmentation-nnUNet.

## Methods

Datasets

We used five independent stroke MRI datasets to evaluate the generalisability of nnU-Net for acute and chronic lesion segmentation. Each dataset included manually delineated lesion masks generated by experts on modality-appropriate sequences, which served as the ground truth. Figure 1A and 1B shows lesion spatial distribution and lesion volume profiles for each dataset.

Acute Stroke Datasets

**(I) Stroke Onset Optimization Project (SOOP)**

The SOOP dataset (N = 1456; [30] comprises acute ischemic stroke cases with high-resolution T1-weighted (T1w) MRI, fluid-attenuated inversion recovery (FLAIR), trace diffusion-weighted imaging (DWI) and apparent diffusion coefficient (ADC) maps. Imaging was acquired within 48 hours of admission. Lesions predominantly involved the middle cerebral artery (MCA) territory, spanning cortical and subcortical regions across both hemispheres. Manual segmentation was performed on combined DWI–ADC images by three trained neuroscientists using MRIcroGL [31]. These DWI–ADC-derived masks served as the reference standard for all acute SOOP model configurations, including FLAIR-only models. In cases with multiple infarcts, all lesions were merged into a single binary mask.

**(II) Ischemic Stroke Lesion Segmentation (ISLES)**

The ISLES dataset (N = 250; [23]) includes DWI, ADC and a mixture of high- and low-resolution FLAIR scans acquired 1–3 weeks post-stroke. Data were collected from multiple sites and scanners, using 1.5T and 3T Philips or Siemens MRI systems. Lesions primarily involved the MCA territory, with approximately one quarter occurring in infratentorial regions and showing mixed hemispheric involvement. Segmentation followed a hybrid pipeline. First, a 3D U-Net algorithm [32] initially trained on DWI images (b = 1000 s/mm²) stemming from one of the centres, was used to generate initial lesion masks. These masks were then manually corrected by medical students using ITK-SNAP [33] or 3D Slicer [34]. The masks were reviewed by neuroradiology residents and verified by senior neuroradiologists. Final masks were generated through joint inspection of DWI, ADC, and FLAIR, and served as a common reference standard for DWI- and FLAIR-based ISLES models. All lesion components were merged into a single binary mask per case.

Chronic Stroke Datasets

**(III) Anatomical Tracings of Lesions After Stroke (ATLAS v2.0)**

ATLAS v2.0 (N = 655; [17,35]) comprises chronic stroke cases (≥180 days post-stroke) with high-resolution T1w MRI and manually delineated lesion masks, pooled across 33 research cohorts. Imaging was acquired on 1.5T and 3T MRI scanners with heterogeneous acquisition parameters. The dataset predominantly consists of single-lesion cases (61.9%), with lesions distributed across cortical and subcortical regions and both hemispheres. Lesions were traced in MRIcron and ITK-SNAP [33] using slice-by-slice manual delineation with optional interpolation for large lesions. Annotators received structured neuroanatomical training and adhered to detailed lesion-tracing guidelines [35]. All segmentations underwent multi-stage quality control, including review by trained raters and consensus resolution with a neuroradiologist. All lesion components were merged into a single binary mask.

### (IV) Aphasia Recovery Cohort (ARC)

The ARC dataset (N = 228; [36]) includes individuals with chronic post-stroke aphasia after left-hemisphere infarcts, scanned on average ≥ 1,138 days post-stroke (SD = 1,327). All T1w scans were acquired on Siemens 3T MRI systems (Trio and later Prisma) across multiple studies, resulting in variability in acquisition parameters. Lesions predominantly affected left MCA territory and were mainly cortical. Although multi-modal imaging (including T2-weighted, DWI and functional MRI) is available for subsets of this cohort, only T1w images were used in the present study to ensure consistency with other chronic datasets. Lesion masks were manually delineated by an expert on the T2-weighted images from the first scanning session.

### (V) Cambridge Cognitive Neuroscience Research Panel (CCNRP)

The CCNRP dataset (N = 204) includes individuals with a broad range of focal brain lesions. Based on clinical diagnosis, approximately 36% of cases reflected stroke-related injury (e.g., ischemic infarct, haemorrhage, aneurysm, or AVM-related events), 53% reflected tumour-related pathology (e.g., meningioma, glioma, cavernoma, and other neoplasms or tumour resection), and the remaining 11% represented other aetiologies, including epilepsy surgery, encephalomalacia, and infections. Lesions exhibited heterogeneous anatomical distributions, spanning cortical and subcortical regions across both hemispheres. All patients were scanned with high-resolution T1w MRI, and lesions were manually delineated by a single expert neurologist. Scans and corresponding manual segmentations are available at OpenNeuro: https://openneuro.org/datasets/ds007212/.

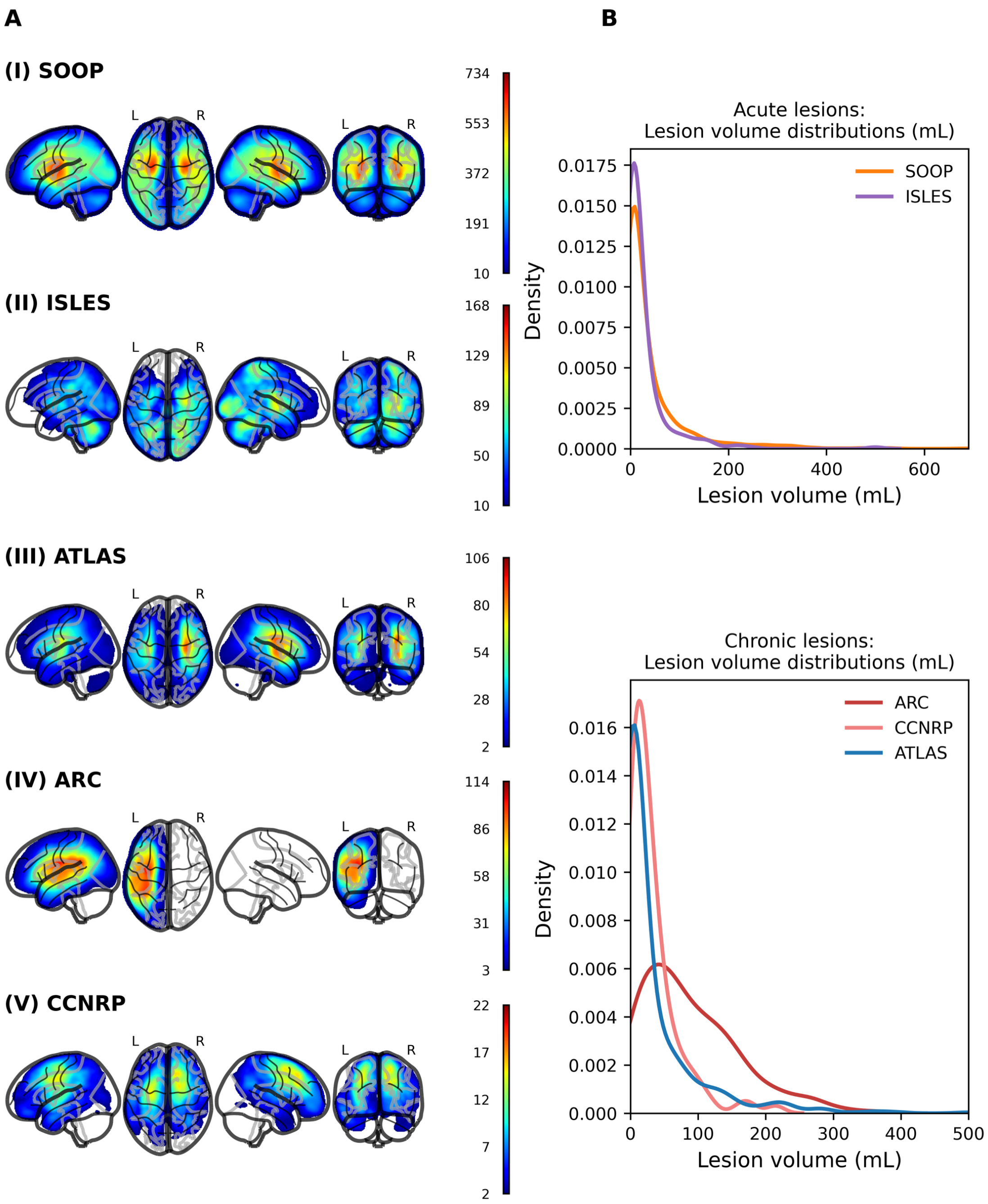

**Figure 1**. Lesion topography and volume distributions across datasets. (A) Glass-brain lesion overlap maps show the number of participants with a lesion at each voxel. Colour bars indicate the number of participants with a lesion (n), and the total sample size is shown for each dataset: SOOP (N = 1,456), ISLES (N = 250), ATLASv2 (N = 655), ARC (N = 228), and CCNRP (N = 204). (B) Kernel density curves show lesion-volume distributions, in millilitres (mL), calculated from binary lesion masks for each dataset. Acute lesion volumes (top) were derived from diffusion-weighted imaging (DWI) masks, whereas chronic lesion volumes (bottom) were derived from T1-weighted (T1w) masks. Abbreviations: SOOP = Stroke Outcome Optimization Project; ISLES = Ischemic Stroke Lesion Segmentation; ATLASv2 = Anatomical Tracings of Lesions After Stroke, version 2; ARC = Aphasia Recovery Cohort; CCNRP = Cambridge Cognitive Neuroscience Research Panel; N = total sample size; n = number of participants with a lesion at a given voxel.

## NnU-Net Architecture

All segmentation models were implemented using the nnU-Net framework (v2.3.1; [28]. Analyses were conducted on a Linux server equipped with five GPUs (3 × NVIDIA A30, 24 GB; 2 ×

NVIDIA L40S, 48 GB). We followed the standard nnU-Net workflow, beginning with automated preprocessing (*nnUNetv2_plan_and_preprocess*), which included metadata extraction, Z-score intensity normalisation, and automatic generation of configuration plans for both the standard full-resolution 3D U-Net (3d_fullres) and a residual-encoder variant (3d_fullres_resenc). To reduce computational burden, we did not train 2D, 3d_lowres, or cascade models. All models were trained using the default nnU-Net settings; a complete list of parameters is provided at https://github.com/AjayHalai/Lesion-Segmentation-nnUNet.

For acute stroke datasets, diffusion inputs consisted of paired DWI and ADC maps, which were always provided together to nnU-Net as a two-channel input (hereafter referred to as DWI). In multimodal experiments (DWI+FLAIR), each sequence was included as a separate image channel. For chronic stroke datasets, models were trained using single-channel T1w images.

Initial model training and internal evaluation (*nnUNetv2_train*) was performed using five-fold cross-validation for each training dataset, using the default loss function (a combination of Dice coefficient and cross-entropy loss) to optimise both spatial overlap and voxel-wise classification. As two architectural configurations were trained per dataset, we used *nnUNetv2_find_best_configuration* to identify the optimal model (either separate or ensemble) based on mean cross-validation Dice scores. The selected model was subsequently applied to independent held-out test datasets (*nnUNetv2_predict*) followed by post-processing (*nnUNetv2_apply_postprocessing*) to remove small, disconnected components and enforce binary masks.

## Evaluation Metrics

Following standard practice in medical image segmentation [37], performance was evaluated using the Dice coefficient and 95th percentile Hausdorff Distance (HD95). The Dice coefficient quantifies spatial overlap between predicted and reference lesion masks, with values ranging from 0 (no overlap) to 1 (perfect overlap). HD95 measures the boundary distance between predicted and reference segmentations and is less sensitive to extreme outliers. It was capped at 256 mm, corresponding to the maximum Euclidean distance within a 1 mm isotropic image grid. In addition, we computed a comprehensive set of voxel-wise and surface-based metrics, including intersection over union (IoU), full Hausdorff distance, mean surface displacement, confusion matrix derived measures, absolute volume difference, and lesion-level F1 score. These additional results are provided in Supplementary Table S1 but are not discussed further here for simplicity.

Statistical Analysis

All analyses were hypothesis-driven and focused on evaluating model performance and generalisability. For clarity, statistical inference is reported primarily on external validation datasets. Group comparisons of Dice and HD95 across experimental conditions were conducted using non-parametric Mann–Whitney U or Wilcoxon signed-rank tests as appropriate (two-sided, $\alpha = .05$). Bonferroni correction was applied where multiple comparisons were performed. Bayes Factor ($BF_{10}$) were calculated using the Pingouin statistical package [38] to quantify the strength of evidence for the alternative versus null hypothesis. Dataset-specific results are provided in the Supplementary Materials Table S1.

**Best-Performing Model Selection**

We first addressed the central question of whether automated models can approach human-level reliability when evaluated on fully independent datasets for each time point (acute and chronic).

For acute stroke, the search space consisted of separate nnU-Net models trained for each modality configuration available within the SOOP and ISLES datasets (DWI, FLAIR, and multi-channel DWI+FLAIR). For chronic stroke, we similarly trained separate models on all T1w images from the ATLASv2, ARC, and CCNRP datasets. Within each time point, all trained models were evaluated on their respective held-out external test datasets, and the model achieving the highest external validation performance was selected as the best acute and chronic model, respectively.

**Effect of Training Set Size (chronic T1w lesions)**

To examine the effect of training size on segmentation, we trained nnU-Net models on three nested subsets of the ATLASv2 dataset (n = 218, 436, 655), randomly split without replacement. Model performance was evaluated on the held-out ARC and CCNRP datasets. Consistent with deep-learning principles, we expected accuracy to improve with larger training set size.

**Effect of Lesion Laterality (chronic T1w lesions)**

The ARC dataset contains exclusively left hemisphere (LH) lesions. Therefore, we examined whether lesion laterality in the training data influences model generalisation. We hypothesised that models trained on laterality-restricted datasets would reduce generalisability to datasets with heterogeneous lesion distributions. To test this, we stratified the ATLASv2 dataset into LH (n = 205), right-hemisphere (RH; n = 224), and bilateral (BL; n = 226) subsets, trained separate models, and evaluated their performance on held-out ATLASv2 and CCNRP datasets.

**Effect of Lesion Volume (chronic T1w lesions)**

We examined the influence of lesion volume (normalised as % intracranial volume) on segmentation performance in two complementary ways. As an outcome measure, we assessed non-linear associations between lesion volume and Dice scores using Spearman correlations across all external validation tests. Additionally, we analysed performance for small lesions using two thresholds ($< 10$ cm$^3$ and $< 20$ cm$^3$) to determine whether volume effects persisted within these restricted size ranges. As a training feature, we first tested generalisation by training models on very small lesions ($< 5$ cm$^3$; n = 319) or larger lesions ($> 5$ cm$^3$ up to ~400 cm$^3$; n = 336) from the ATLASv2 dataset. Each model was then evaluated on the alternate subset (small → large; large → small), as well as on the held-out ARC and CCNRP datasets. Secondly, given that ATLASv2 is heavily skewed toward smaller lesions, we tested whether a broader lesion-volume distribution improves performance. To do this, we constructed a volume-matched subset of ATLASv2 (n = 204) that approximated the lesion-volume distribution of CCNRP. A model trained on this volume-matched ATLASv2 subset was evaluated on CCNRP, and the previously trained CCNRP model was likewise evaluated on the matched ATLASv2 subset, allowing us to assess generalisation in both directions. Matching either dataset to ARC was not feasible due to the substantially larger and left-lateralised lesions present in the ARC dataset.

**Effect of Image Quality (chronic T1w lesions)**

Finally, we assessed whether differences in MRI quality contributed to variability in performance. Image quality metrics were extracted from T1w images using MRIQC [39] for the three chronic datasets. MRIQC provides a broad set of quantitative measures that can be grouped into: 1) noise-related measures (e.g., contrast- or signal-to-noise ratio; CNR and SNR, respectively), 2) information-theoretic measures (e.g., foreground-background energy ratio and energy-focus criterion), 3) artifact-related measures (e.g., bias field intensity, artifact signal in air space, and white matter intensity), and 4) structural measures (e.g., blurriness, residual partial volume effects, intracranial volume, tissue probability maps overlapping with MNI template). A full description of these metrics is provided in the original MRIQC publication.

Metrics with very low variance across participants (< 0.01 after standardisation) were excluded, and redundant measures capturing overlapping information (e.g., axis-specific smoothness metrics fwhm_x, fwhm_y, fwhm_z) were removed. All metrics were recoded such that higher values reflected better image quality. To reduce dimensionality and summarise latent qualities, we applied principal component analysis (PCA) with varimax rotation, retaining components with eigenvalues > 1. Variables with absolute loadings > 0.60 were used to interpret each component. Component scores were compared across datasets to evaluate systematic MRI quality differences.

To directly evaluate whether MRI image quality was associated with segmentation performance, MRIQC-derived factor scores from each individual test scan were correlated with the corresponding Dice score obtained for that same case during external validation using Spearman correlations. Partial correlations controlling for lesion volume were additionally performed to assess whether MRI quality effects were independent of lesion size. Bayesian correlations were used to quantify evidence strength.

## Results

Below, we report model performance results structured around the key hypotheses outlined above. Unless otherwise specified, external validation results are reported by pooling all cases across the relevant held-out test datasets for each comparison.

**Best Performing Models: Acute and Chronic Lesions**

For acute lesions, the SOOP-trained DWI+FLAIR model achieved the highest segmentation accuracy (median Dice = .777; median HD95 = 2.45 mm) and statistically outperformed the SOOP-trained DWI-only model (median Dice = .774; Wilcoxon $W$ = 11,464.5, $p$ = .004). However, the absolute improvement was small (Δ Dice = .003, $BF_{10}$ = 0.96). Accordingly, the DWI-only model may represent a practical clinical option given it is most common with minimal performance drop. Across all modality configurations, DWI-only trained models consistently outperformed FLAIR models (all $p < .001$; all BF10 > $10^{63}$; Δ Dice range = .29–.56). Details of all models are shown in Supplementary Materials Table S1.

For chronic stroke, the best-performing model was the ATLASv2 model, which achieved a median Dice of .77 (HD95 = 4.64) on externally validation datasets. This model significantly outperformed both the CCNRP-trained model (median Dice = .63; median HD95 = 11.92 mm) and the ARC-trained model (median Dice = .002; median HD95 = 67.57 mm) (all $p < .001$; all $BF_{10} > 1000$). Dataset-specific external validation results are provided in Supplementary Table S1.

Taken together, these results demonstrate that nnU-Net can achieve performance on heterogeneous, independent datasets that falls within the range of published inter-rater reliability (Dice ≈ .72–.84; [14,15,23]).

Upon inspecting individual cases we found that segmentation errors, reflected by low Dice and/or high HD95, arose from both model limitations and annotation inaccuracies. In some instances, low Dice reflected clear model failures, such as missing lesions. In other cases, discrepancies were attributable to limitations in the ground truth annotations, including mislabelled voxels outside the brain, under-segmentation, or inconsistently defined lesion boundaries. These examples indicate that evaluation metrics may reflect a combination of model and human annotation errors. Representative examples are shown in Figure 2.

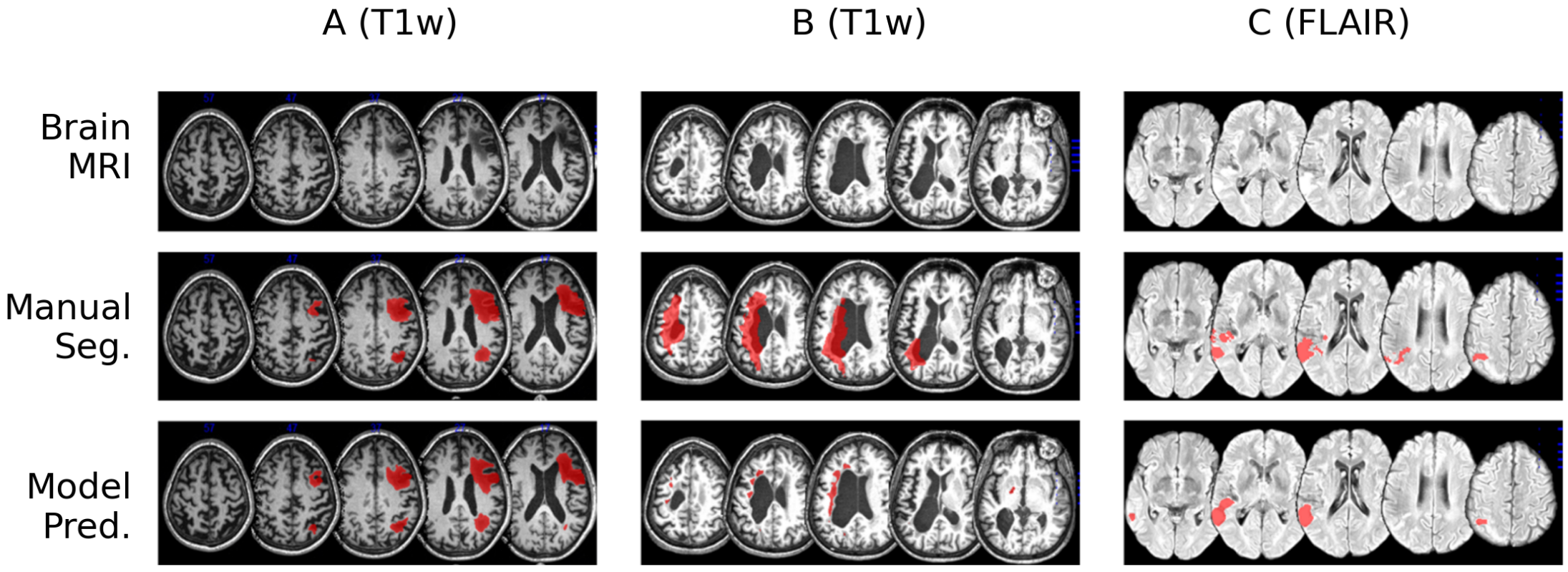

**Figure 2**. Segmentation examples of stroke lesions across datasets. Each panel shows one illustrative case (N = 1 per panel). For each case, axial images in Montreal Neurological Institute (MNI) space are shown at the top, manual lesion masks (ground truth; red) are overlaid in the middle, and model predictions (red) are overlaid at the bottom. (A) Chronic T1-weighted (T1w) example: a model trained on ATLASv2 (training N = 655) was evaluated on the Cambridge Cognitive Neuroscience Research Panel (CCNRP; N = 204), showing a large cortical lesion with close agreement between prediction and manual mask (Dice = 0.90; HD95 = 2.25 mm). (B) Chronic T1w example: the same ATLASv2-trained model (training N = 655) was evaluated on the Aphasia Recovery Cohort (ARC; N = 228), showing a large lesion with poor overlap between prediction and manual mask (Dice = 0.30; HD95 = 21 mm). Here, the discrepancy arises primarily from differences in lesion-definition conventions, for example whether enlarged ventricles or periventricular cerebrospinal fluid should be considered part of the lesion. (C) Acute fluid-attenuated inversion recovery (FLAIR) example: a model trained on Stroke Outcome Optimization Project (SOOP) FLAIR images (training N = 1,456) was evaluated on Ischemic Stroke Lesion Segmentation (ISLES) FLAIR images (N = 250), showing a cortical lesion with moderate agreement between prediction and manual mask (Dice = 0.55; HD95 = 7.3 mm). This example highlights the challenges of FLAIR, where other hyperintense structures, including white-matter hyperintensities, oedema, or venous artefacts, can resemble stroke lesions and contribute to human and model ambiguity. Abbreviations: ATLASv2 = Anatomical Tracings of Lesions After Stroke, version 2; Dice = Dice similarity coefficient; HD95 = 95th-percentile Hausdorff distance; mm = millimetres; N = sample size.

## Effect of Training Set Size (chronic lesions; T1w)

Training set size had a modest yet significant effect on segmentation performance across external validation datasets. Dice scores increased significantly from the small (n = 218; median = .739) to the medium-sized model (n = 436; median = .762; $W = 20{,}220$, $z = -10.22$, $p < .001$, $BF_{10} > 1000$). Consistent with this trend, the large model also significantly outperformed the medium model (N = 655; median = .774; $W = 34{,}050$, $z = -4.90$, $p < .001$, $BF_{10} > 1000$). However, the magnitude of improvement was diminished beyond a few hundred training cases (small to medium Δ Dice = .023 vs. medium to large Δ Dice = .012), suggesting it is not necessary to have larger training cases. Dice score distributions are shown in Figure 3, with summary metrics in Table 1. For simplicity, results are pooled across both external validation datasets; dataset-specific results are provided in Supplementary Materials Table S1.

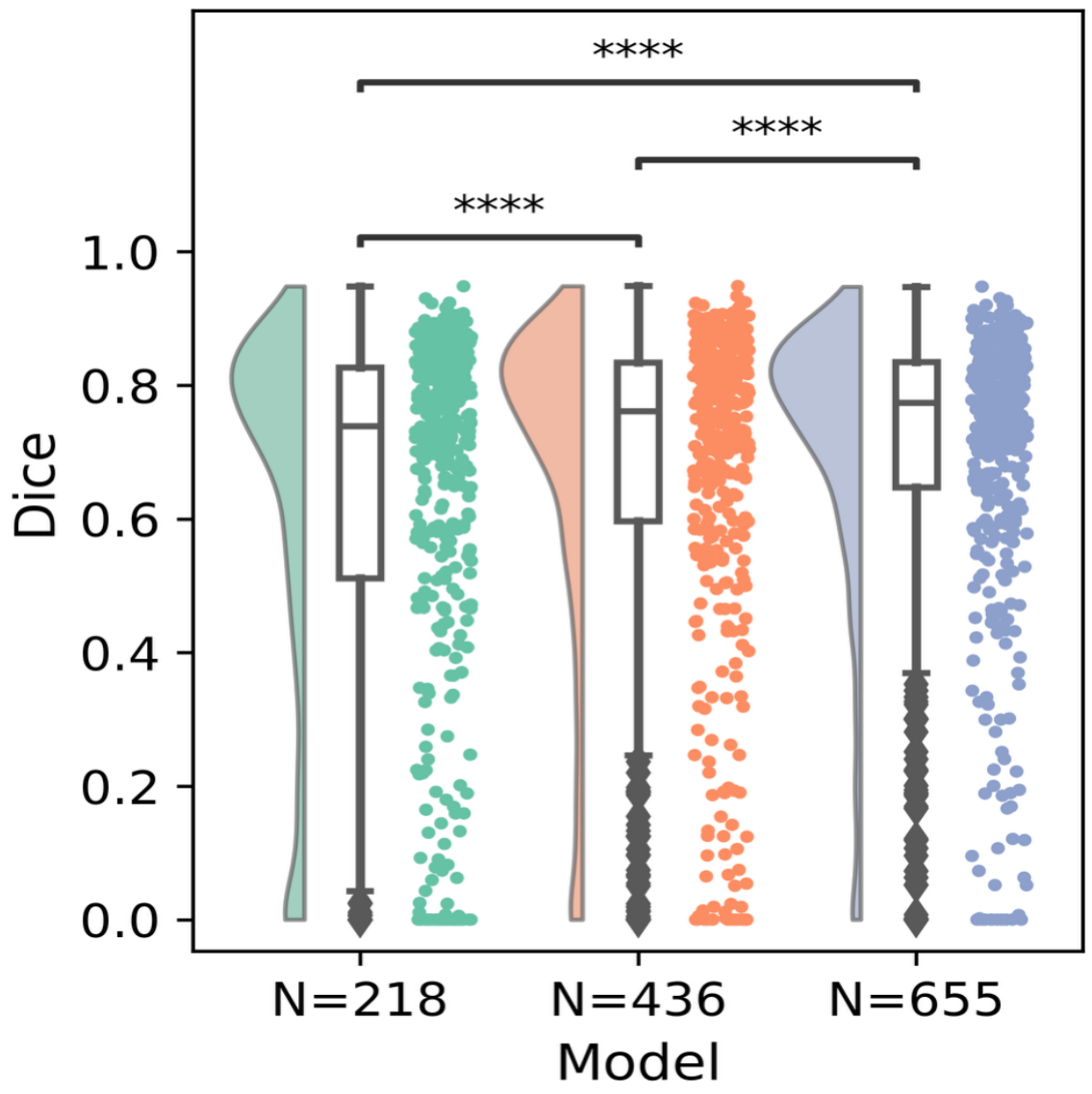

**Figure 3**. External-validation Dice scores for models trained on increasingly large ATLASv2 subsets. The x-axis indicates ATLASv2 training subset size (N = 218, 436, and 655); the y-axis shows Dice coefficients on pooled external ARC and CCNRP test sets (N = 432; ARC, N = 228; CCNRP, N = 204). Violin plots show score distributions and boxplots show the median and interquartile range. Each dot represents one external validation case and its Dice score. Pairwise two-sided Wilcoxon signed-rank tests with Bonferroni correction showed differences between models trained on 218 versus 436 scans (W = 18,300, Padj = $4.95 \times 10^{-24}$), 218 versus 655 scans (W = 17,000, Padj = $4.04 \times 10^{-27}$), and 436 versus 655 scans (W = 32,140, Padj = $1.91 \times 10^{-6}$). **** Padj < .0001. Abbreviations: ATLASv2 = Anatomical Tracings of Lesions After Stroke, version 2; ARC = Aphasia Recovery Cohort; CCNRP = Cambridge Cognitive Neuroscience Research Panel; Dice = Dice similarity coefficient; N = sample size; W = Wilcoxon signed-rank test statistic; Padj = Bonferroni-adjusted P value.

**Table 1** Segmentation performance (median Dice and HD95) of nnU-Net models trained on chronic stroke lesions.

| Testing Target | Training set | External test set | Dice (Median, Range) | HD95 (Median, Range) |
|---|---|---|---|---|
| **Training size** | ATLASv2 Small (N=218) | ARC, CCNRP | .739 (0–.95) | 6.2 (1, 256) |
| | ATLASv2 Medium (N=436) | ARC, CCNRP | .762 (0–.95) | 5.1 (1, 256) |
| | ATLASv2 Large (N=655) | ARC, CCNRP | .774 (0–.95) | 4.6 (1, 256) |
| **Dataset origin** | ARC (N = 228) | CCNRP, ATLASv2 | .002 (0–.92) | 67.57 (1, 256) |
| | ATLASv2 (N = 655) | ARC, CCNRP | .774 (0–.95) | 4.64 (1, 256) |
| | CCNRP (N = 204) | ATLASv2, ARC | .626 (0–.92) | 11.92 (1, 256) |
| **Laterality** | ATLASv2 LH (N = 205) | ARC, CCNRP | .744 (0, .94) | 5.96 (1, 256) |
| | ATLASv2 RH (N = 224) | ARC, CCNRP | .750 (0, .94) | 5.43 (1, 256) |
| | ATLASv2 BL (N = 226) | ARC, CCNRP | .740 (0, .95) | 5.38 (1, 256) |
| **Lesion volume** | ATLASv2 small (N = 319) | ATLAS big | .074 (0–.85) | 44.8 (1.4–256) |
| | | ARC, CCNRP | .005 (0-.82) | 55.1 (2.2–256) |
| | ATLASv2 big (N = 236) | ATLAS small | .545 (0–.91) | 22.1 (1–256) |
| | | ARC, CCNRP | .773 (0-.95) | 4.3 (1–256) |

LH = left hemisphere; RH = right hemisphere; BL = bilateral lesions; HD95 = 95th percentile Hausdorff Distance.

### Lesion Laterality Effects (chronic lesions; T1w)

The ARC trained model showed poor performance on external datasets (see Table 1), raising the possibility that this reflected its restriction to left hemisphere lesions. To test this, we trained separate models on laterality-restricted subsets of ATLASv2 (LH: $n = 205$; RH: $n = 224$; BL: $n = 226$), and compared their performance to the full ATLASv2 trained model that contained all cases (N = 655). Lesion volumes did not differ between LH and RH subsets ($p > .05$), although both unilateral groups contained significantly larger lesions than the bilateral subgroup (LH vs BL: U = 6,109, $p < .001$ and RH vs BL: U = 7,723, $p < .001$). All laterality-restricted models performed similarly with median Dice ranging from .74–.75, and Bayesian analyses supported the absence of meaningful differences between LH, RH, and BL models (all $BF_{10} < 0.2$; Figure 4; Table 1). One exception was a small but significant reduction in performance for the LH model relative to the full ATLASv2 model (Dice = .74 vs. .77; U = 83,390, p = .041, $BF_{10}$ = 57.23). However, the absolute effect size was small (Δ Dice ≈ .03), suggesting limited practical impact on model generalisation. Full dataset-specific results are provided in Supplementary Materials Table S1.

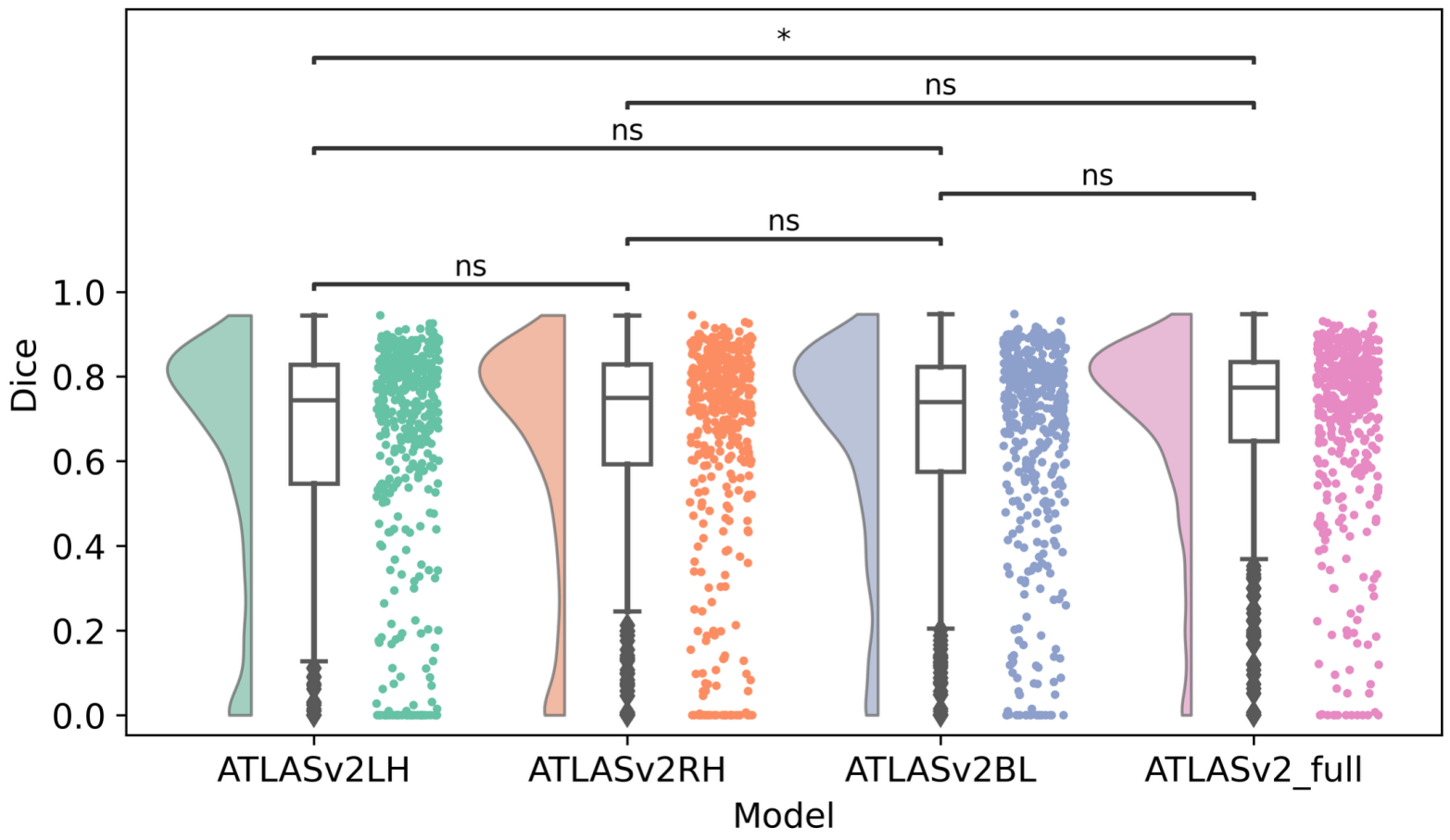

**Figure 4.** Dice score distributions for models trained on laterality-specific ATLASv2 subsets (LH, N = 205; RH, N = 224; BL, N = 226) and the full ATLASv2 dataset (N = 655). The x-axis indicates the training subset and the y-axis shows Dice coefficients on the pooled external ARC and CCNRP test sets (N = 432; ARC, N = 228; CCNRP, N = 204). Boxplots show the median and interquartile range; each dot represents one external validation case and its Dice score. Pairwise two-sided Mann-Whitney U tests with Bonferroni correction showed no significant differences between LH and RH (U = 91,840, Padj = 1.000), LH and BL (U = 91,530, Padj = 1.000), RH and BL (U = 93,120, Padj = 1.000), BL and the full dataset (U = 84,570, Padj = .103), or RH and the full dataset (U = 84,880, Padj = .129). Dice scores were lower for the LH-trained than full-dataset model (U = 83,390, Padj = .041). * Padj < .05. Abbreviations: ATLASv2 = Anatomical Tracings of Lesions After Stroke, version 2; LH = left hemisphere; RH = right hemisphere; BL = bilateral; ARC = Aphasia Recovery Cohort; CCNRP = Cambridge Cognitive Neuroscience Research Panel; Dice = Dice similarity coefficient; N = sample size; U = Mann-Whitney U statistic; Padj = Bonferroni-adjusted P value.

## Lesion Volume Effects (chronic lesions; T1w)

We next examined the influence of lesion volume on segmentation accuracy, considering lesion volume as both (i) a property of the test set affecting performance on individual cases, and (ii) a characteristic of the training set, influencing generalisation.

Test-Set Effects: Smaller Lesions are Harder to Segment

Spearman rank correlations revealed strong positive associations between lesion volume and Dice score across all models and external datasets ($r_s(202\text{-}653) = .49–.74$, all $p < .001$), indicating that larger lesions were segmented more accurately. An illustrative example for the ATLASv2 model is shown in Figure 5.

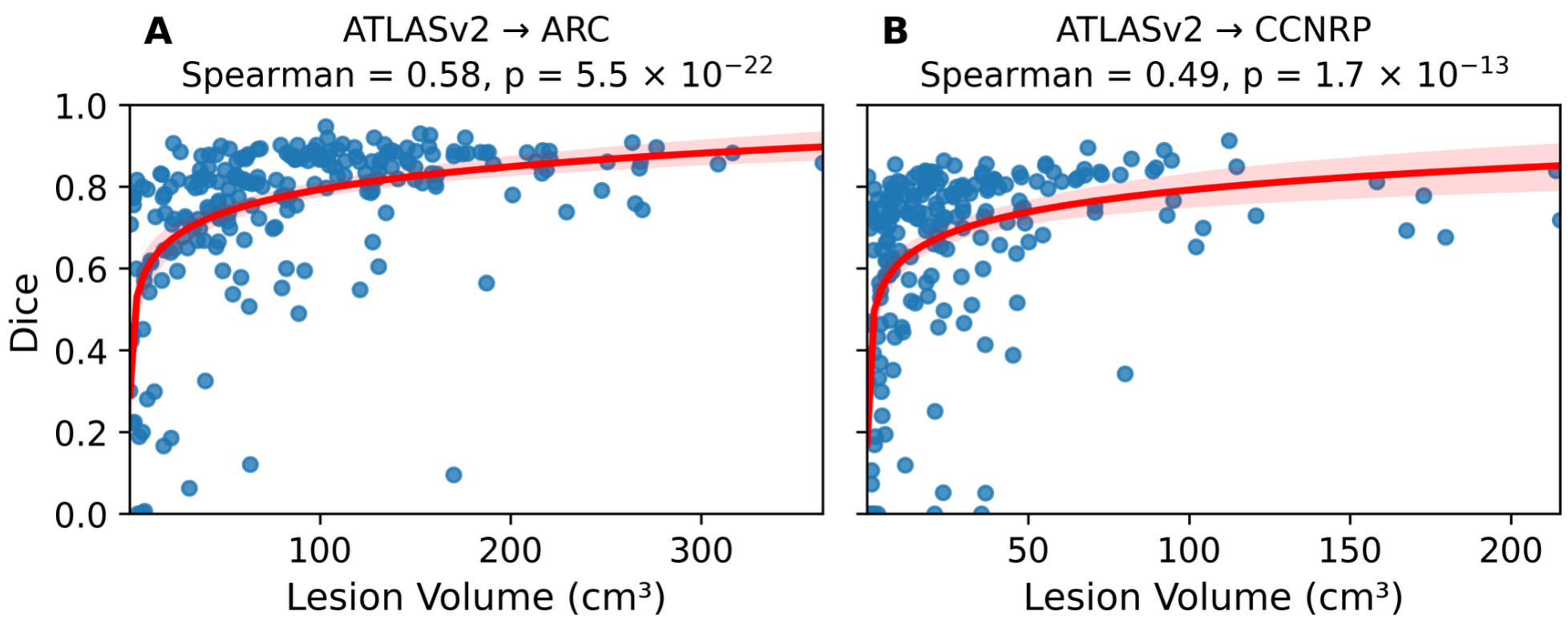

**Figure 5**. Spearman's correlation between lesion volume and segmentation accuracy for a model trained on ATLASv2 (training N = 655). Scatterplots show the association between lesion volume (x-axis) and Dice coefficient (y-axis) for predictions on ARC (A; N = 228) and CCNRP (B; N = 204). Each blue dot represents one external test case and its corresponding lesion mask. Red curves depict nonlinear trend fits with 95% confidence intervals (shaded areas). Spearman rank correlation coefficients and corresponding P values are reported above each plot. Abbreviations: ATLASv2 = Anatomical Tracings of Lesions After Stroke, version 2; ARC = Aphasia Recovery Cohort; CCNRP = Cambridge Cognitive Neuroscience Research Panel; Dice = Dice similarity coefficient; $cm^3$ = cubic centimetres; N = sample size.

When analyses were restricted to small lesions ($\leq 20$ $cm^3$), segmentation difficulty persisted. Lesions with poor segmentation accuracy (Dice ≤ .6) were significantly smaller than those with better segmentation accuracy (Dice > .6; U = 91–10,940, all $p < .05$, $BF_{10} > 10$), indicating that lesion volume continued to influence performance even within this restricted range. In contrast, within the subset of very small lesions (< 10 $cm^3$), lesion volume no longer distinguished between well- and poorly segmented cases across all datasets (all $p > .05$, $BF_{10} < 1$). We plotted the lesion overlap map of the well- and poorly segmented lesions from the small lesion (<10 $cm^3$) but did not find obvious spatial patterns, although the poorly segmented lesions appeared more spatially dispersed (Supplementary Material Figure S1).

### Training Set Effects: Small-Lesion Trained Models Generalise Poorly

To assess whether training on very small lesions improves generalisation to similar cases, we compared two models trained on volume-stratified ATLASv2 subsets: a small-lesion model (median= 1.14 $cm^3$; n = 319) and a large-lesion model (median = 34.01 $cm^3$; n = 336).

Internal validation revealed substantially poorer performance for the small-lesion model compared to the large-lesion model (Dice = .57 vs. .82; U = 20,780, $p < .001$, $BF_{10} > 1000$). While the large-lesion model generalised moderately to held-out small lesions (Dice = .54), the small-lesion model failed to generalise to larger lesions (Dice = .07, $BF_{10} > 1000$). This asymmetry was amplified in external datasets. For ARC, the large-lesion model significantly outperformed the small-lesion model (Dice = .82 vs. .01; U = 1,616, $p < .001$, $BF_{10} > 1000$), with a comparable pattern observed for CCNRP (Dice = .74 vs. .00; U = 2,052, $p < .001$, $BF_{10} > 1000$).

Volume-Matched Analysis: Interactions of Training and Test Sets

The chronic datasets differed substantially in their lesion-volume distribution; therefore, we conducted a volume-matched analysis to isolate the contribution of lesion volume to cross-dataset generalisation. Lesion volume differed significantly across datasets (ARC > CCNRP > ATLASv2; all $p < .001$), motivating the construction of a subset of ATLASv2 (n = 204) whose lesion-volume distribution was statistically equivalent to CCNRP (3.95% vs. 4.08%; U = 20,812, $p > .05$).

Prior to matching, cross-dataset performance was asymmetric: the CCNRP-trained model performed poorly on the full ATLASv2 dataset (Dice = .50), whereas an ATLASv2-trained model generalised well to CCNRP (Dice = .73; asymmetry: U = 10,940, $p < .001$, $BF_{10} > 1000$). When evaluation was restricted to the volume-matched ATLASv2 subset, this asymmetry was eliminated (CCNRP→ATLASv2_matched: Dice = .63; ATLASv2_matched→CCNRP: Dice = .66; U = 20,140, $p > .05$, $BF_{10} = 0.12$). This reflects a significant improvement of the CCNRP-trained model relative to its performance on the full ATLASv2 dataset (U = 53,960, $p < .001$, $BF_{10} > 1000$), consistent with the matched subset containing larger, intrinsically easier to segment lesions. Conversely, the volume-matched ATLASv2-trained model exhibited slightly reduced performance on CCNRP relative to the full ATLASv2 model (U = 24,682, $p = .001$, $BF_{10} = 32.01$), likely reflecting subtle differences in lesion-volume coverage between the matched and full ATLASv2 training sets despite comparable overall performance

## Image Quality Effect (chronic lesions; T1w)

A varimax-rotated PCA of MRIQC-derived quality metrics yielded a five-component solution explaining 83.1% of the total variance. Full factor loadings are provided in Supplementary Table S2 and definitions of all MRIQC metrics included in the factor analysis are provided in Supplementary Material Table S3. Briefly, Factor 1 reflected signal integrity, Factor 2 tissue-segmentation precision, Factor 3 spatial resolution and tissue contrast, Factor 4 signal consistency, and Factor 5 background contrast-to-noise characteristics.

A Kruskal–Wallis test revealed significant differences between datasets across all five components (H = 7.48–397.88, all $p < .05$). Follow-up Mann–Whitney U tests with Bonferroni correction showed that ARC scored significantly lower than both ATLAS and CCNRP on Factors 1, 2 and 4 (all p's < .001, all $BF_{10} > 1000$), corresponding to reduced SNR, poorer tissue boundary definition, and less homogeneous signal. ARC also scored lower than CCNRP on Factor 5 (U = 26,281, p = .05, $BF_{10} > 1000$), reflecting weaker contrast-to-noise ratio and noisy background. In contrast, ARC scored significantly higher than other datasets on Factor 3 (U = 4,578–93,509, all $p < .001$, $BF_{10} > 1000$), a dimension characterised by complex positive and negative loadings which was less straightforward to interpret. Standardised factor profiles across datasets are visualised in Figure 6.

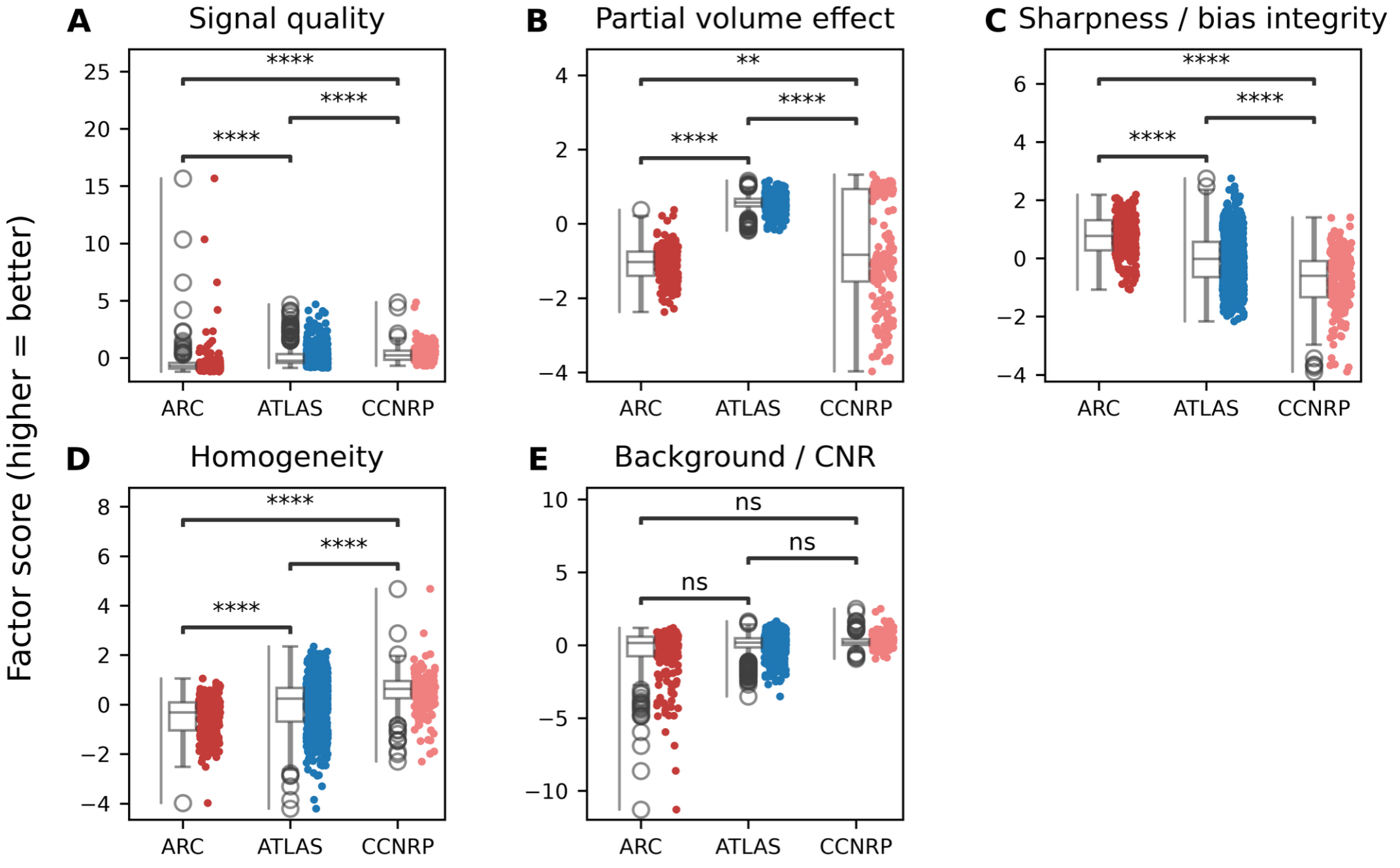

**Figure 6.** Standardized MRIQC factor scores by dataset. (A) Signal quality (factor 1; F1); (B) partial volume effect (factor 2; F2); (C) sharpness/bias integrity (factor 3; F3); (D) homogeneity (factor 4; F4); and (E) background/contrast-to-noise ratio (factor 5; F5). Factor scores were derived from a varimax-rotated principal component analysis of MRIQC metrics. Metrics were oriented before analysis so that higher factor scores indicate better image quality. Boxes represent the median and interquartile range. Each coloured point represents the standardized factor score from one participant's T1-weighted scan (ARC, N = 228; ATLASv2, N = 655; CCNRP, N = 204). Group differences were assessed using Kruskal-Wallis tests followed by pairwise Mann-Whitney U tests with Bonferroni correction. Significant pairwise comparisons are indicated ($P < .05$, *$P < .01$, **$P < .001$, ***$P < .0001$; full test statistics are reported in the results. Abbreviations: MRIQC = MRI Quality Control; ARC = Aphasia Recovery Cohort; ATLASv2 = Anatomical Tracings of Lesions After Stroke, version 2; CCNRP = Cambridge Cognitive Neuroscience Research Panel; N = sample size.

Spearman correlations further revealed significant associations between MRI quality factors and Dice scores for Factors 1, 3 and 4 (Spearman's $\rho = -.28$ to .29, all $p < .001$), with Bayesian analyses providing moderate-to-extreme evidence for these relationships ($BF_{10}$ = 396 to $8.41\times10^9$). After controlling for lesion volume using partial Spearman correlations, only Factor 3 remained significantly associated with Dice scores ($\rho = .17$, $p < .001$, $BF_{10} = 3382.80$). Importantly, these analyses demonstrate that image quality of the test scans was associated with segmentation accuracy, although they do not directly test whether lower-quality training datasets impair cross-dataset generalisation.

## Discussion

Reliable and scalable automated lesion segmentation is a key target for stroke neuroscience, with direct implications for research, clinical prognosis, and personalised care [3,11,12,40]. In this study, we systematically evaluated the performance and generalisability of the nnU-Net framework across multiple heterogeneous acute and chronic stroke MRI datasets. The best-performing models demonstrated robust external, out-of-sample generalisation for both acute (median = .78) and chronic (median = .73-.81) cases. These values fall within, and in some cases exceed, reported inter-rater reliability for human expert manual annotations [16,17], highlighting the promise of automated approaches for large-scale research and as a foundation for future clinical deployment. However, the results demonstrated that high accuracy was possible under specific conditions and some model configurations clearly failed (e.g., acute FLAIR, chronic ARC, and chronic small lesions only). These findings highlight both the promise and the boundary conditions of current automated lesion segmentation approaches. Below, we situate these findings within the existing literature, discuss their implications, and outline challenges and opportunities for advancing toward clinically robust lesion segmentation.

### Acute Lesion Models: DWI vs. FLAIR

For acute stroke, models trained on DWI consistently generalised well to independent datasets, while combining FLAIR with DWI produced only minimal gains and, in some cases, reduced accuracy relative to DWI alone (Δ Dice = .003); Bayesian analyses provided little evidence for a meaningful advantage of the multimodal model. This comparison requires cautious interpretation because the available reference masks were not independently delineated within each input modality. In SOOP, masks were manually defined on DWI–ADC images and were subsequently used to train and evaluate both DWI- and FLAIR-based models. In ISLES, the initial masks were generated using a DWI-based algorithm and the final masks were reviewed using DWI, ADC and FLAIR. Thus, although ISLES annotations incorporated FLAIR information, the reference standards were more aligned with DWI than with FLAIR. This methodological asymmetry is expected to favour DWI-based similarity scores; accordingly, the present analyses cannot establish that DWI is intrinsically superior to FLAIR for acute lesion segmentation. Nevertheless, the lower performance of FLAIR-based models, including during internal validation and in ISLES, where final annotations incorporated all three sequences, suggests that label provenance may not be the sole contributor to the observed differences.

Differences in acquisition timing and the presence of competing FLAIR hyperintensities may also have contributed to the lower FLAIR performance observed in the present analysis. In the hyperacute phase, restricted diffusion is evident on DWI (high signal with low ADC), while FLAIR signal often remains normal, producing the classic DWI–FLAIR mismatch. Over subsequent days, lesions typically appear hyperintense on both DWI and FLAIR, with continued ADC reduction. By the subacute stage (1–3 weeks), DWI signal declines and ADC values pseudo-normalise, while FLAIR hyperintensity persists. Beyond this period, DWI becomes more variable with elevated ADC values [23]. Differences in acquisition timing between datasets (SOOP <48 hours; ISLES 1-3 weeks) possibly contributed to variability in lesion appearance and influenced cross-data generalisation.

The lower performance of FLAIR-based models during internal validation is also consistent with the influence of features that make FLAIR lesion delineation challenging. In older populations, white matter hyperintensities and other chronic vascular changes can obscure lesion boundaries or mimic acute infarction[42–44]. While FLAIR remains clinically valuable for estimating stroke onset (e.g., [45]), future work could explicitly model white matter hyperintensities as a separate class to disentangle chronic vascular pathology from acute ischaemic injury.

More broadly, these modality-specific differences highlight a central challenge: stroke lesions are highly dynamic, with physiological and morphological changes unfolding across time. Visibility varies not only between sequences but also within the same sequence across stages [46]. Models trained solely on acute-phase scans are therefore unlikely to generalise to chronic-stage lesions without explicitly accounting for lesion evolution. Next-generation approaches should aim to capture these temporal dynamics, enabling longitudinal and multimodal predictions.

**Dataset Size During Training Matters**

For chronic stroke, increasing training set size was associated with systematic yet modest performance gains, with the largest improvements observed when moving from small to medium datasets and more modest gains thereafter. Although Bayes factor provided strong evidence for incremental improvements with increasing training size, the absolute Dice gains became progressively smaller, suggesting diminishing practical benefit beyond a few hundred training cases. This pattern aligns with broader findings in medical imaging, where performance improves with additional data until gains plateau [47]. Practically, our findings suggest that moderately sized but diverse datasets (~400–500 cases) can already support robust nnU-Net models for chronic T1w lesion segmentation. These diminishing returns likely reflect redundancy in homogeneous data rather than limits of model capacity, underscoring the importance of diversity over sheer sample size [48].

**Lesion laterality does not constrain generalisation**

Despite marked differences in lesion laterality across datasets, laterality alone did not substantially limit generalisation. Models trained on left, right or bilateral lesions achieved comparable performance, and Bayes factor further supported the absence of meaningful differences between laterality-restricted models. These findings align with prior work suggesting that lesion volume, shape and anatomical dispersion are more critical determinants of segmentation accuracy than laterality [49,50]. The poor external performance of ARC-trained models therefore likely reflects other factors discussed below.

**Lesion volume as key determinant of accuracy and generalisability**

Lesion volume emerged as one of the strongest and most consistent predictors of segmentation performance [21,51]. Across all models and datasets, larger lesions were segmented more accurately, with strong positive correlations between lesion volume and Dice scores. Smaller lesions, particularly those below 20 $cm^3$, were consistently more difficult to segment, consistent with previous reports [3,50,52]. This likely reflects subtle boundaries, greater anatomical variability, and lower tissue contrast – factors that also reduce inter-rater reliability in manual tracing [14].

Importantly, lesion volume also shaped generalisability through its representation in the training data. Models trained on restricted lesion-volume distributions, particularly predominantly small lesions, generalised poorly outside their respective volume ranges, whereas models trained on broad volume distributions showed more symmetric cross-dataset performance. Volume-matching analyses further demonstrated that apparent generalisation asymmetries largely disappeared once lesion distributions were aligned, suggesting that lesion-volume distribution itself may be a driver of external validation performance.

Interestingly, models trained with larger lesions (> 5 $cm^3$) performed similarly to the full ATLAS dataset, despite having a third of training data. We noted above that training set size only hinted a modest performance increase beyond a few hundred cases; however, this result indicates that breadth of lesion distribution within the training data could be very important. In the current model, the large lesions data comprised of a board range (5- 496.656 $cm^3$), and future studies could investigate this systematically to determine if range of data and the lesion size interact during training.

**Does Image Quality Matter?**

We provide tentative evidence that MRI quality contributed to variation in generalisation. Using MRIQC-derived metrics summarised via PCA, we found that the ARC dataset exhibited systematically lower image quality across multiple dimensions, including SNR, tissue boundary definition, and signal homogeneity. Bayes factor scores provided extremely strong evidence for systematic MRI quality differences between datasets across multiple latent quality dimensions. These factors likely limited robust feature learning during training, despite reasonable internal accuracy [53,54]. Interestingly, and potentially importantly for future clinical translation to quality-varying scans, ARC cases were segmented relatively well by models trained on higher-quality datasets, suggesting that training on diverse, high-quality data may confer some robustness to lower-quality scans at inference. These findings align with broader machine learning literature showing that training with low-quality data can impair generalisation [55–57]. Our results could not formally demonstrate that the image quality of the training data itself was linked to segmentation performance. We attempted to split the chronic datasets we had into high/low image scans but were unable to control for other characteristics (e.g., lesion size and distribution); therefore, we suggest future studies would need to collate an appropriate dataset to test this hypothesis directly.

## Towards Clinical Translation

Although this study focused on MRI-based segmentation, clinical translation ultimately requires consideration of real-world imaging workflows, particularly in acute stroke where CT remains the primary imaging modality. While CT offers lower soft-tissue contrast than MRI, segmentation may be enhanced using anatomical priors, tissue probability maps, and using multimodal CT–MRI datasets acquired within the same time window [3,20,58–60]. Our findings motivate future work extending automated lesion segmentation to CT and multimodal clinical data.

## Limitations

Several limitations should be acknowledged. Automated models are constrained by the quality and consistency of human-generated ground truth, which is known to be variable, particularly for small lesions [14]. In some cases, model outputs appeared more anatomically coherent than the reference masks, highlighting ambiguity in the annotations rather than definitive model error. Future work may benefit from consensus-based labels, multi-rater probabilistic maps, uncertainty-aware learning approaches, or integrating vascular territory information[61]. In addition, although Dice is widely used for segmentation evaluation, it is particularly sensitive to small voxel mismatches in very small lesions. While multiple complementary performance metrics are provided in the Supplementary Materials, future work should evaluate whether volume-normalised measures may provide a more appropriate assessment of segmentation quality in this setting. Annotation protocols also introduce modality-specific bias. In both acute datasets, lesions were delineated primarily with reference to DWI, potentially disadvantaging FLAIR-based models. Sequence-specific annotations or multi-label approaches may better capture modality-dependent lesion appearance. In addition, chronic analyses in the current study were restricted to T1-weighted imaging because additional modalities were not consistently available across the chronic datasets. Future work should evaluate whether multimodal approaches incorporating sequences such as FLAIR can further improve chronic lesion segmentation. Finally, we did not exhaustively explore advanced multimodal architectures or pretraining strategies, which may leverage complementary information beyond the models tested here.

## Conclusion

This study provides a comprehensive evaluation of automated stroke lesion segmentation using nnU-Net across acute and chronic MRI datasets. We demonstrated robust external generalisation, with performance commensurate with human reliability, while identifying lesion volume, training data quality and dataset diversity as key determinants of accuracy. Acute segmentation may benefit more from DWI images, but future research is required to test FLAIR segmentation trained and tested within modality. Chronic segmentation improves with moderate training set size before plateauing. Together, these findings offer practical guidance for dataset curation and model development and outline a roadmap towards more robust, generalisable, and clinically relevant lesion segmentation.

## Data Availability

All trained models and code are openly shared to promote transparency, reproducibility, and community benchmarking: (https://github.com/AjayHalai/Lesion-Segmentation-nnUNet). The data used in the study are open-source and can found at the following websites: SOOP (https://openneuro.org/datasets/ds004889/versions/1.1.2); ISLES (https://isles22.grand-challenge.org/); ATLASv2 (https://atlas.grand-challenge.org/); ARC

(https://openneuro.org/datasets/ds004884); CCNRP (https://openneuro.org/datasets/ds007212/).

## Acknowledgements

We gratefully acknowledge the teams responsible for the SOOP, ISLES, ATLASv2, ARC, and CCNRP datasets for making these resources openly available to the research community.

## Funding

This work was supported by an Israel Scholarship Education Foundation award (ISEF; T.T.); Blavatnik Fellowship award (T.T.); Medical Research Council Career Development Award (MR/V031481/1 to A.D.H.); Medical Research Council programme grant (MR/R023883/1 to M.A.L-R.) and Medical Research Council intramural funding (MC_UU_00005/18 to M.A.L-R.).

## Competing interests

The authors report no competing interests.

## Supplementary material

Supplementary material is available online.

# Figure Legends

Figure 1. Lesion topography and volume distributions across datasets. (A) Glass-brain lesion overlap maps show the number of participants with a lesion at each voxel. Colour bars indicate the number of participants with a lesion (n), and the total sample size is shown for each dataset: SOOP (N = 1,456), ISLES (N = 250), ATLASv2 (N = 655), ARC (N = 228), and CCNRP (N = 204). (B) Kernel density curves show lesion-volume distributions, in millilitres (mL), calculated from binary lesion masks for each dataset. Acute lesion volumes (top) were derived from diffusion-weighted imaging (DWI) masks, whereas chronic lesion volumes (bottom) were derived from T1-weighted (T1w) masks. Abbreviations: SOOP = Stroke Outcome Optimization Project; ISLES = Ischemic Stroke Lesion Segmentation; ATLASv2 = Anatomical Tracings of Lesions After Stroke, version 2; ARC = Aphasia Recovery Cohort; CCNRP = Cambridge Cognitive Neuroscience Research Panel; N = total sample size; n = number of participants with a lesion at a given voxel.

**Alt text:** Composite figure with two panels. (A) Five glass-brain maps display lesion-overlap patterns for the acute SOOP and ISLES datasets and the chronic ATLASv2, ARC, and CCNRP datasets. Colour scales indicate the number of participants with a lesion at each voxel. (B) Line plots show lesion-volume distributions for acute datasets (top) and chronic datasets (bottom). In both panels, lesion volumes are concentrated at lower values with long right-tailed distributions toward larger lesions.

Figure 2. Segmentation examples of stroke lesions across datasets. Each panel shows one illustrative case (N = 1 per panel). For each case, axial images in Montreal Neurological Institute (MNI) space are shown at the top, manual lesion masks (ground truth; red) are overlaid in the middle, and model predictions (red) are overlaid at the bottom. (A) Chronic T1-weighted (T1w) example: a model trained on ATLASv2 (training N = 655) was evaluated on the Cambridge Cognitive Neuroscience Research Panel (CCNRP; N = 204), showing a large cortical lesion with close agreement between prediction and manual mask (Dice = 0.90; HD95 = 2.25 mm). (B) Chronic T1w example: the same ATLASv2-trained model (training N = 655) was evaluated on the Aphasia Recovery Cohort (ARC; N = 228), showing a large lesion with poor overlap between prediction and manual mask (Dice = 0.30; HD95 = 21 mm). Here, the discrepancy arises primarily from differences in lesion-definition conventions, for example whether enlarged ventricles or periventricular cerebrospinal fluid should be considered part of the lesion. (C) Acute fluid-attenuated inversion recovery (FLAIR) example: a model trained on Stroke Outcome Optimization Project (SOOP) FLAIR images (training N = 1,456) was evaluated on Ischemic Stroke Lesion Segmentation (ISLES) FLAIR images (N = 250), showing a cortical lesion with moderate agreement between prediction and manual mask (Dice = 0.55; HD95 = 7.3 mm). This example highlights the challenges of FLAIR, where other hyperintense structures, including white-matter hyperintensities, oedema, or venous artefacts, can resemble stroke lesions and contribute to human and model ambiguity. Abbreviations: ATLASv2 = Anatomical Tracings of Lesions After Stroke, version 2; Dice = Dice similarity coefficient; HD95 = 95th-percentile Hausdorff distance; mm = millimetres; N = sample size.

**Alt text:** Three columns of axial brain MRI images compare manual lesion masks with nnU-Net predictions. In each column, the top row shows the anatomical image, the middle row shows the manual lesion mask in red, and the bottom row shows the predicted mask in red. (A) A chronic cortical lesion is closely matched by the prediction.

(B) A chronic lesion shows poor agreement, particularly around enlarged ventricles. (C) An acute FLAIR image shows moderate agreement between the predicted and manual cortical lesion masks.

Figure 3. External-validation Dice scores for models trained on increasingly large ATLASv2 subsets. The x-axis indicates ATLASv2 training subset size (N = 218, 436, and 655); the y-axis shows Dice coefficients on pooled external ARC and CCNRP test sets (N = 432; ARC, N = 228; CCNRP, N = 204). Violin plots show score distributions and boxplots show the median and interquartile range. Each dot represents one external validation case and its Dice score. Pairwise two-sided Wilcoxon signed-rank tests with Bonferroni correction showed differences between models trained on 218 versus 436 scans (W = 18,300, Padj = $4.95 \times 10^{-24}$), 218 versus 655 scans (W = 17,000, Padj = $4.04 \times 10^{-27}$), and 436 versus 655 scans (W = 32,140, Padj = $1.91 \times 10^{-6}$). **** Padj < .0001. Abbreviations: ATLASv2 = Anatomical Tracings of Lesions After Stroke, version 2; ARC = Aphasia Recovery Cohort; CCNRP = Cambridge Cognitive Neuroscience Research Panel; Dice = Dice similarity coefficient; N = sample size; W = Wilcoxon signed-rank test statistic; Padj = Bonferroni-adjusted P value.

**Alt text:** Violin, box, and dot plots compare Dice scores for models trained on 218, 436, and 655 ATLASv2 scans and evaluated on 432 pooled external cases. Dice-score distributions shift upward as training-set size increases. Brackets above the plots indicate significant Bonferroni-corrected pairwise Wilcoxon signed-rank tests between all three training-set sizes.

Figure 4. Dice score distributions for models trained on laterality-specific ATLASv2 subsets (LH, N = 205; RH, N = 224; BL, N = 226) and the full ATLASv2 dataset (N = 655). The x-axis indicates the training subset and the y-axis shows Dice coefficients on the pooled external ARC and CCNRP test sets (N = 432; ARC, N = 228; CCNRP, N = 204). Boxplots show the median and interquartile range; each dot represents one external validation case and its Dice score. Pairwise two-sided Mann-Whitney U tests with Bonferroni correction showed no significant differences between LH and RH (U = 91,840, Padj = 1.000), LH and BL (U = 91,530, Padj = 1.000), RH and BL (U = 93,120, Padj = 1.000), BL and the full dataset (U = 84,570, Padj = .103), or RH and the full dataset (U = 84,880, Padj = .129). Dice scores were lower for the LH-trained than full-dataset model (U = 83,390, Padj = .041). * Padj < .05. Abbreviations: ATLASv2 = Anatomical Tracings of Lesions After Stroke, version 2; LH = left hemisphere; RH = right hemisphere; BL = bilateral; ARC = Aphasia Recovery Cohort; CCNRP = Cambridge Cognitive Neuroscience Research Panel; Dice = Dice similarity coefficient; N = sample size; U = Mann-Whitney U statistic; Padj = Bonferroni-adjusted P value.

**Alt text:** Violin, box, and dot plots compare Dice scores for models trained on left-hemisphere, right-hemisphere, bilateral, and full ATLASv2 datasets, evaluated on 432 pooled external cases. Score distributions are broadly similar across training subsets. A bracket indicates that the left-hemisphere-trained model has modestly lower Dice scores than the full-dataset model; no other pairwise comparisons are significant after Bonferroni correction.

Figure 5. Spearman's correlation between lesion volume and segmentation accuracy for a model trained on ATLASv2 (training N = 655). Scatterplots show the association between lesion volume (x-axis) and Dice coefficient (y-axis) for predictions on ARC (A; N = 228) and CCNRP (B; N = 204). Each blue dot represents one external test case and its corresponding lesion mask. Red curves depict nonlinear trend fits with 95% confidence intervals (shaded areas). Spearman rank correlation coefficients and corresponding P values are reported above each plot. Abbreviations: ATLASv2 = Anatomical Tracings of Lesions After Stroke, version 2; ARC = Aphasia Recovery Cohort; CCNRP = Cambridge Cognitive Neuroscience Research Panel; Dice = Dice similarity coefficient; $cm^3$ = cubic centimetres; N = sample size.

**Alt text:** Two scatterplots show a positive association between lesion volume and Dice score for external ARC and CCNRP test cases. Blue dots represent individual cases. In both panels, red fitted curves rise rapidly at small lesion volumes and then plateau, indicating lower segmentation accuracy for very small lesions and higher accuracy for larger lesions. Shaded bands show 95% confidence intervals.

**Figure 6.** Standardized MRIQC factor scores by dataset. (A) Signal quality (factor 1; F1); (B) partial volume effect (factor 2; F2); (C) sharpness/bias integrity (factor 3; F3); (D) homogeneity (factor 4; F4); and (E) background/contrast-to-noise ratio (factor 5; F5). Factor scores were derived from a varimax-rotated principal component analysis of MRIQC metrics. Metrics were oriented before analysis so that higher factor scores indicate better image quality. Boxes represent the median and interquartile range. Each coloured point represents the standardized factor score from one participant's T1-weighted scan (ARC, N = 228; ATLASv2, N = 655; CCNRP, N = 204). Group differences were assessed using Kruskal-Wallis tests followed by pairwise Mann-Whitney U tests with Bonferroni correction. Significant pairwise comparisons are indicated ($P < .05$, $^*P < .01$, $^{**}P < .001$, $^{***}P < .0001$; full test statistics are reported in the results. Abbreviations: MRIQC = MRI Quality Control; ARC = Aphasia Recovery Cohort; ATLASv2 = Anatomical Tracings of Lesions After Stroke, version 2; CCNRP = Cambridge Cognitive Neuroscience Research Panel; N = sample size.